\documentclass[
reprint,
 amsmath,amssymb,
 aps,prl,showpacs
]{revtex4-1}

\usepackage{graphicx}% Include figure files
\usepackage{dcolumn}% Align table columns on decimal point
\usepackage{bm}% bold math
\usepackage{hyperref}% add hypertext capabilities
\begin{document}

\preprint{APS/123-QED}

\title{Piezoelectric electromechanical coupling\\ in nanomechanical resonators with two-dimensional electron gas}% Force line breaks with \\

\author{A. A. Shevyrin}
 \email{shevandrey@isp.nsc.ru}
\author{A. G. Pogosov}
\author{A. K. Bakarov}
\author{A. A. Shklyaev}
\affiliation{%
 Rzhanov Institute of Semiconductor Physics\\
13, Lavrentyev ave., Novosibirsk, Russia, 630090}
\affiliation{%
Novosibirsk State University\\
2, Pirogov str., Novosibirsk, Russia, 630090}

\date{\today}% It is always \today, today,
             %  but any date may be explicitly specified

\begin{abstract}
The electrical response of two-dimensional electron gas to vibrations of a nanomechanical cantilever containing it is studied. Vibrations of perpendicularly oriented cantilevers are experimentally shown to change oppositely the conductivity near their bases. This indicates the piezoelectric nature of electromechanical coupling. A physical model is developed, which quantitatively explains the experiment. It shows that the main origin of the conductivity change is a rapid change in the mechanical stress on the boundary between suspended and non-suspended areas, rather than the stress itself.
\end{abstract}

\pacs{85.85.+j, 63.22.-m, 73.50.Dn, 77.65.Ly}% PACS, the Physics and Astronomy
                             % Classification Scheme.
%\keywords{Suggested keywords}%Use showkeys class option if keyword
                              %display desired
\maketitle

%\tableofcontents

%\section{\label{sec:level1}Introduction\protect}

Most of the currently studied low-dimensional electron systems are fabricated from a two-dimensional electron gas (2DEG) embedded in a semiconductor bulk. A classical example of such a system is a 2DEG in GaAs/AlGaAs heterostructures. However, selective etching of a sacrificial layer (often called surface nanomachining) gives an opportunity to create also a 2DEG embedded in a thin membrane freely suspended over a substrate \cite{blick98}. The nanostructures fabricated from such membranes are mechanically moveable with their movement affecting electron transport and conductivity \cite{tang02}. Such electromechanical coupling gives an opportunity to probe mechanical motion at the nano-scale and it could be used to study various interesting mechanical phenomena, such as "phonon lasing" \cite{mahboob14} and the quantum-limited motion of an artificially made object \cite{cleland10}. Moreover, it opens up new prospects for studying non-trivial transport phenomena in 2DEG under unusual conditions, namely, in the presence of additional mechanical degrees of freedom, and for creating nanoelectromechanical systems (NEMS). For example, papers \cite{cleland02,yamaguchi13} show that the electron transport through a quantum point contact placed on a micromechanical resonator is sensitive to mechanical vibrations. Papers \cite{tang02,shev15,shev13} demonstrate that diffusive conductive channels in 2DEG can also be used as nanoelectromechanical transducers.

The two fundamental key points arising in the context of NEMS are the physical mechanisms underlying actuation and transduction of the nanomechanical motion. The question about the actuation in NEMS with 2DEG is addressed elsewhere \cite{shev15}, while, in the present paper, we focus on the transduction mechanism. Most of the papers considering GaAs/AlGaAs-based suspended systems contain a proposal that a 2DEG embedded in a resonator is sensitive to its vibrations due to the change in the density of a 2DEG that screens the piezoelectrically induced bound charge \cite{beck, tang02, cleland02, yamaguchi13}. However, there is a lack of experimental evidence for this hypothesis.

In the present paper, we experimentally demonstrate that the dominant physical mechanism making a 2DEG sensitive to NEMS mechanical vibrations is associated with the piezoelectric effect and show the sensitivity magnitude. We propose also a physical model giving an independent estimate for the value of electron density change consistent with the experiment. According to the model, the local change in the 2DEG conductivity is determined mainly by spatial deviations of the mechanical stress tensor, rather than by the stress itself.

The piezoelectric effect is essentially anisotropic \cite{masmanidis2007,vopilkin} and, in a GaAs crystal, identical mechanical stresses in the perpendicular directions  [110]  and [$\bar{1}$10]  induce opposite electrical polarizations. The central idea of the experiment is to check whether the change in the conductivity of a 2DEG, contained in two identically vibrating cantilevers oriented in the considered directions, reflects  such anisotropy and, thus, to test the hypothesis about the piezoelectric nature of electromechanical coupling. 

The experimental samples are fabricated from the GaAs/AlGaAs heterostructure described in detail in \cite{shev15}. The heterostructure contains a 166 nm-thick stack of layers grown by means of molecular-beam epitaxy above a 400 nm-thick Al$_{0.8}$Ga$_{0.2}$As sacrificial layer, which, in turn, resides on the [001]-oriented GaAs substrate. The stack contains two Al$_{0.33}$Ga$_{0.67}$As layers surrounding the 13 nm-thick GaAs layer with a 2DEG. Also, the stack contains a 10 nm-thick GaAs top cap layer. The 2DEG has electron density of $n=6.7\times{10^{11}}$ cm$^{-2}$ and the mobility of $\mu=1.2\times{10^6}$ cm$^2$V$^{-1}$s$^{-1}$. The samples lateral geometry is defined in a single electron-beam lithography step followed by anisotropic plasmachemical etching in $\mathrm{BCl_3}$. The samples suspension is performed by means of selective wet etching in a 1:100 hydrofluoric acid water solution.

\begin{figure}[b]
\includegraphics{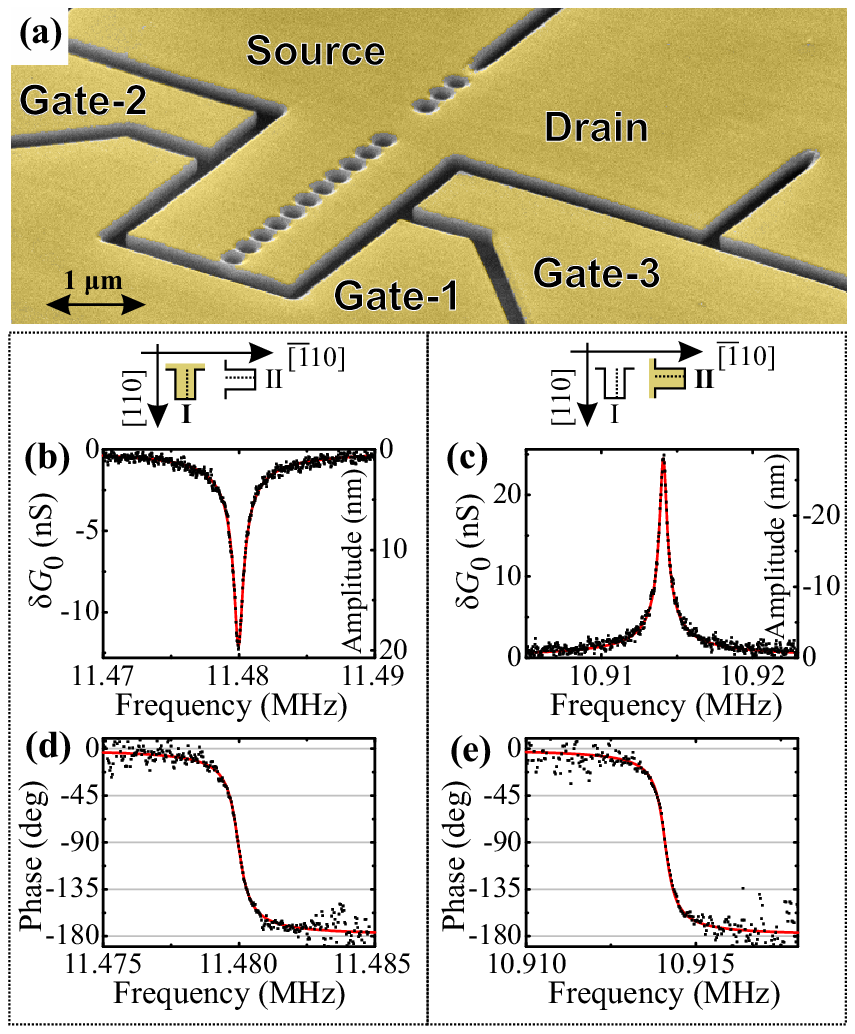}
\caption{\label{fig1} (a) False-color scanning-electron-microscope image of a cantilever containing a two-dimensional electron gas. The areas between the holes in the cantilever are non-conductive due to edge depletion, except for the one constriction where the inter-hole distance is enlarged. (b-e) Signed amplitudes (b,c) and phases (d,e) of the current flowing through the cantilevers oriented along [110] (b,d) and [$\bar{1}$10] (c,e) crystallographic directions.}
\end{figure}

Each experimental chip contains two identical nanomechanical cantilevers oriented in perpendicular directions ([110]  and [$\bar{1}$10]). Hereinafter, we will refer to them as Cantilever-I and Cantilever-II, respectively. The cantilevers are $L=3\ \mathrm{\mu m}$ long,  $W=2\ \mathrm{\mu m}$ wide and $t=166\ \mathrm{nm}$ thick (see Fig. \ref{fig1}(a)). There is a series of holes placed on a single longitudinal line on each of the cantilevers. The distance between the holes is small enough for the regions between them being non-conductive due to edge depletion, except for one spacing near the cantilever base, where the distance is enlarged up to 600 nm. Thus, the line of holes electrically separates the 2DEG into two areas - the source and the drain (individual for each cantilever) - connected via a single constriction. Each cantilever is equipped with three side gates with one of them (Gate-1) surrounding the free end and two others (Gate-2 and Gate-3) placed near the base.

The conductance $G$ response to the cantilevers vibrations is measured using the heterodyne down-mixing technique \cite{bargatin} applied in the following way. Let a cantilever perform flexural vibrations at the fundamental mode. Assuming that the vibrations are small, consider the cantilever as a driven linear oscillator, which motion is described by the following equation:
\begin{equation}
\ddot \xi+ \frac {\Omega_0}{Q}\dot\xi+\Omega_0^2\xi=\frac{F_0\cos{\Omega t}}{m}\label{Newton},
\end{equation}
where $\xi$ is displacement of the cantilever free end, $\Omega_0$ is resonant frequency, $Q$ is the quality factor, $F_0$ and $\Omega$ are the amplitude and the frequency of the effective driving force, and $m$ is the cantilever effective mass. Write the solution of (\ref{Newton}) as
\begin{equation}
\xi=\xi_0(\Omega)\cos{[\Omega t+\varphi(\Omega)]}\label{solut}.
\end{equation}
Since the vibrations are small, consider only the linear conductance response to the vibrations:
\begin{equation}
G=G_0+\frac{dG}{d\xi}\xi=G_0+\delta G_0\cos{[\Omega t+\varphi (\Omega)]}\label{response}
\end{equation}

To transform this high-frequency response into a low-frequency electrical signal, we apply a voltage
\begin{equation}
V_{SD}=V_0\cos{(\Omega-\omega)t}\label{VSD}
\end{equation}
between the source and the drain. Here $\omega=2\pi\times 25\mathrm{kHz}\ll\Omega,\Omega_0$. Source-drain current $I=GV_{SD}$ has two components at the high $2\Omega-\omega$ and the low $\omega$ heterodyne frequencies. The low-frequency component which we measure in the experiment is
\begin{equation}
I_\omega=I_0(\Omega)\cos{[\omega t+\varphi(\Omega)]},\label{current}
\end{equation}
where $I_0(\Omega)=V_0\delta G_0(\Omega)/2$. Thus, measuring amplitude $I_0(\Omega)$ and the phase of this low-frequency current, we obtain the amplitude $\delta G_0(\Omega)=2I_0(\Omega)/V_0$ and the phase $\varphi(\Omega)$ of the high-frequency conductance response to vibrations. At the same time, the heterodyne down-mixing eliminates the known difficulties inherent to the measurements at the high driving frequency \cite{bargatin}.

The cantilevers vibrations are driven at the fundamental flexural mode perpendicularly to the surface using the electrostatic (capacitive) actuation scheme. For this purpose, we apply a voltage $V_G$ to Gate-1 (see Fig. \ref{fig1}(a)), satisfying condition $\lvert V_G\rvert\gg\lvert V_{SD}\rvert$. Then the effective driving force is
\begin{equation}
F=C^\prime V_G^2/2,\label{force}
\end{equation}
where factor $C^\prime$ is proportional to the derivative of the gate-cantilever capacitance on the free end displacement $\xi$ (we assume $\xi>0$ if the cantilever is bent up in the direction from the substrate). As shown in \cite{shev15},  $C^\prime$ can be estimated as
\begin{equation}
C^\prime\approx-0.39\varepsilon_0WL/d_0^2,
\end{equation}
where $\varepsilon_0$ is the vacuum dielectric constant and $d_0=400\ \mathrm{nm}$ is the distance between the cantilever and the underlying substrate.

The applied voltage $V_G$ is a sum of a dc component $V_{DC}$ and an ac component:
\begin{equation}
V_G=V_{DC}+V_{AC}\cos{(\Omega-\omega)t}\times[1+\cos{\omega t}].\label{VG}
\end{equation}

The ac component is a high-frequency signal (proportional to $V_{SD}$) amplitude-modulated by low frequency $\omega$. The modulated signal has three components: one at carrier frequency $\Omega-\omega$ and two at sidebands $\Omega-2\omega$ and $\Omega\approx\Omega_0$. The effective force acting at the frequency $\Omega$ obtained from  (\ref{force}) and (\ref{VG}) has the amplitude
\begin{equation}
F_0=C^\prime V_{DC}V_{AC}/2.\label{famplit}
\end{equation}

Since, as we show later, $\omega\gg\Omega_0/Q$, the other frequency components of the effective force are far from the resonance and their influence can be neglected.

We use a Tektronix AFG3252C two-channel arbitrary function generator and appropriate attenuators to apply the gate and source-drain voltages. The modulating signal is applied to the generator input from an SR5210 lock-in amplifier. The lock-in amplifier is also used to measure the amplitude $I_0(\Omega)$ and the phase $\varphi(\Omega)$ of the current flowing through the cantilever at modulation frequency $\omega$. The phase is measured with respect to the modulating signal. During the experiment, the samples are placed in a vacuum tube and cooled down to liquid helium temperature 4.2 K.

Figs. \ref{fig1}(b) and 1(c) show the signed amplitude $\delta G_0(\Omega)$ of the conductance response measured as a function of driving frequency $\Omega/2\pi$ for the Cantielever-I and Cantilever-II, respectively. Phase $\varphi(\Omega)$ is shown in Figs. \ref{fig1}(d) and 2(e). The curves are obtained at $V_{DC}=2\ \mathrm{V}$, $V_{AC}=50\ \mathrm{mV}$ and $V_0=6.25\ \mathrm{mV}$.

The main difference between the data obtained for the perpendicularly oriented cantilevers is the sign of their electrical response to vibrations. This difference could equivalently be shown by a $180^\circ$ phase shift, but we use the signed amplitude for clarity. The amplitude-frequency dependence has a Lorentzian form and agrees with (\ref{Newton}) and (\ref{response}), as well as the measured phase-frequency dependence. The solid red lines in Figs. \ref{fig1}(b-e) show the corresponding fits to the experimental data. The resonant frequencies $\Omega_0/2\pi$ extracted from the fits are 11.455 MHz and 10.889 MHz for the Cantilever-I and Cantilever-II, respectively. Quality factors $Q$ are 18000 and 23400. The measured resonant frequencies agree with the rough estimate \cite{landau}
$\Omega_0/2\pi=0.16t\sqrt{E/\rho}/L^2\approx 14.8\ \mathrm{MHz}$, which can be obtained for the first flexural mode of thin cantilever vibrations perpendicular to the surface. Here $E=121\ \mathrm{GPa}$ is the [110] Young modulus of $\mathrm{Al_{0.33}Ga_{0.67}As}$ \cite{adachi} and $\rho=4800\ \mathrm{kg/m^3}$ is the mass density. Some discrepancy can be explained by the fact that the cantilevers width is not small in comparison with their length, as well as by their non-uniformity and by the etching undercut.

Consider the factors influencing the sign of $\delta G_0(\Omega_0)$. This resonant amplitude of the conductance change is proportional to the vibrations amplitude which, in turn, can be expressed from (\ref{Newton}) and (\ref{famplit}) as
\begin{equation}
\xi_0(\Omega_0)=\frac{F_0Q}{m\Omega_0^2}=\frac{C^\prime V_{DC}V_{AC}Q}{2m\Omega_0^2}.\label{amplit}
\end{equation}
Thus, $\delta G_0(\Omega_0)$ should be proportional to dc gate voltage $V_{DC}$ and should change the sign with the negation of $V_{DC}$, if the vibrations are electrostatically driven. Fig. \ref{fig2}(a) shows the experimentally measured $\delta G_0(\Omega_0)$ dependence on $V_{DC}$. The shown data confirm the predicted behavior, with $\delta G_0(\Omega_0)$ having opposite signs for the two cantilevers in all the range of $V_{DC}$. Since the sign of $\xi_0(\Omega_0)$ does not depend on crystallographic orientation, the observed negation of $\delta G_0(\Omega_0)$, according to (\ref{response}), shows that $dG/d\xi$ has the opposite signs for the cantilevers oriented in [110] and $[\bar{1}10]$ directions. This anisotropy can be considered as an indicative of the piezoelectric nature of the mechanical vibrations influence on the conductance. To obtain an additional quantitative confirmation of this hypothesis and to reveal the details of piezoelectric response, we have compared independent experimental and numerical estimates of the relative sensitivity to the vibrations $(1/G_0)(dG/d\xi)$. This value is the proportionality factor between the relative conductance change and the vibrations amplitude:
\begin{figure}[b]
\includegraphics{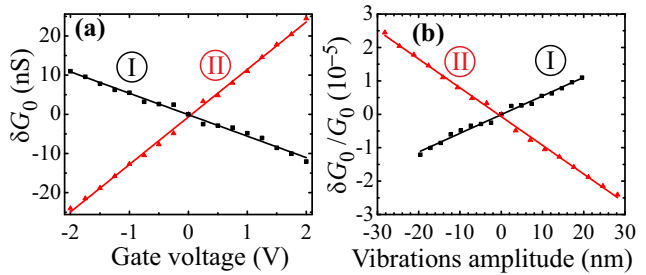}
\caption{\label{fig2} a) The signed amplitudes of the conductance change induced by vibrations have the opposite signs for the Cantilever-I and the Cantilever-II in all the gate voltage range. b) Relative amplitude of the conductance change as a function of the estimated vibrations amplitude.}
\end{figure}
\begin{equation}
\frac{\delta G_0(\Omega_0)}{G_0}=\frac{1}{G_0}\frac{dG}{d\xi}\xi_0(\Omega_0).\label{relat}
\end{equation}

The $\delta G_0(\Omega_0)/G_0$ dependence on $\xi_0(\Omega_0)$ estimated using (\ref{amplit}) is shown in Fig. \ref{fig2}(b). The cantilevers resistance $1/G_0$ measured independently equals $1\ \mathrm{k\Omega}$. A cantilever effective mass appearing in (\ref{amplit}) is estimated as $m=0.24\rho tWL$ \cite{shev15}. The data shown in Fig. \ref{fig2}(b) give the desired values of $(1/G_0)(dG/d\xi)$ equal to $5.6\times 10^{-4}\ \mathrm{\mu m^{-1}}$ and $-8.6\times 10^{-4}\ \mathrm{\mu m^{-1}}$ for the Cantilever-I and the Cantilever-II, respectively. The rest of the paper is devoted to a physical model giving an independent estimate for the relative conductance sensitivity.

\begin{figure}[b]
\includegraphics{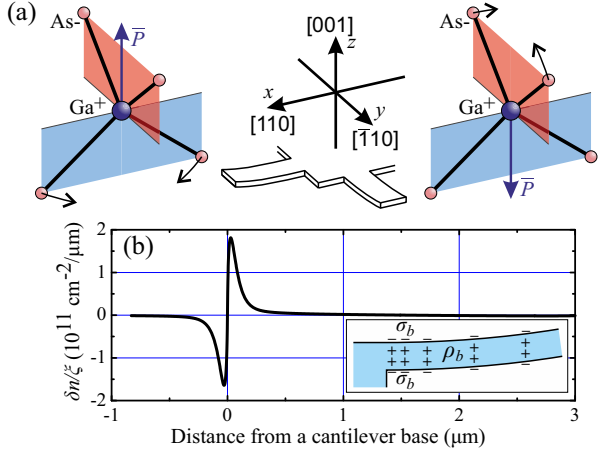}
\caption{\label{fig3} (a) Simplified picture showing the origin of the piezoelectric response anisotropy. (b) Spatial dependence of the estimated change in electron density $\delta n$ per unit displacement $\xi$ of the cantilever free end. The shown dependence corresponds to the $[110]$-oriented cantilever and should be negated for the $[\bar{1}10]$-oriented cantilever.}
\end{figure}

To estimate roughly the mechanical stress, we use the Euler-Bernoulli beam theory and neglect the fact that cantilevers width $W=2\ \mathrm{\mu m}$ is not much less than their length $L=3\ \mathrm{\mu m}$. Then the shape of the first flexural mode is \cite{landau,shev15}
\begin{equation}
U=\xi\times [A(\cos{kl}-\cosh{kl})+B(\sin{kl}-\sinh{kl})]\label{bending}
\end{equation}
where $l$ is the distance from a cantilever base, $k\approx1.875/L$, $A\approx-0.5$ and $B\approx0.367$. Introduce a coordinate system with x-, y- and z-axes coinciding with [110], $[\bar{1}10]$ and [001] crystallographic directions, respectively (see Fig. \ref{fig3} (a)). Then the six-dimensional vectors \cite{nye} describing the mechanical stress (in Voigt notation) are
\begin{eqnarray}
\sigma_i^I=\begin{pmatrix}\sigma&0&0&0&0&0\end{pmatrix}^T,\nonumber\\
\sigma_i^{II}=\begin{pmatrix}0&\sigma&0&0&0&0\end{pmatrix}^T\label{sixvect}
\end{eqnarray}
for the Cantilever-I and the Cantilever-II. Here
\begin{equation}
\sigma=-Ezd^2U/dl^2,\label{sigma}
\end{equation}
where $z=0$ corresponds to the cantilever neutral plane.
The piezoelectric effect leads to electrical polarization $P_i=d_{ij}\sigma_j$, where $d_{ij}$ is the following piezoelectric matrix \cite{nye}:
\begin{equation}
d_{ij}=d_{14}\begin{pmatrix}0&0&0&0&1&0\\0&0&0&-1&0&0\\1&-1&0&0&0&0\end{pmatrix}.\label{matrix}
\end{equation}
Here $d_{14}=-3.04\ \mathrm{pm/V}$ is the $\mathrm{Al_{0.33}Ga_{0.67}As}$ piezoelectric constant \cite{adachi}. The polarization is non-uniform and induces the volume bound charge with the density independent on $z$:
\begin{equation}
\rho_b=-\mathrm{div}{\bm{P}}=\mp d_{14}\frac{d\sigma}{dz}=\mp d_{14}E\frac{d^2U}{dl^2}\label{vcharge},
\end{equation}
 where signs "$-$" and "$+$" correspond to the Cantilever-I and the Cantilever-II, respectively (see Fig. \ref{fig3} (a)). The volume charge is compensated by the bound charge
\begin{equation}
\sigma_b=-\rho_bt/2\label{scharge}
\end{equation}
on the upper and lower surfaces of the cantilever.

Let the electrical potential created by the bound charge be $\delta\phi_\mathrm{ext}$. The 2DEG responds to this external influence with a change in electron density $\delta n$, which, in turn, leads to the change in chemical potential $\delta n\pi\hbar^2/m$ and to the change in electrical potential $\delta\phi_\mathrm{resp}$, such that the electrochemical potential remains zero:
\begin{equation}
-e(\delta\phi_\mathrm{ext}+\delta\phi_\mathrm{resp})+\delta n\frac{\pi\hbar^2}{m^*}=0.{\label{elchem}}
\end{equation}
Here $e$ and $m^*$ are negated electron charge and effective mass in GaAs. To estimate $\delta n$, we neglect the last term in (\ref{elchem}), use the model of pure electrostatic screening \cite{larkin} and consider the 2DEG as having constant electrical potential $\delta\phi_\mathrm{ext}+\delta\phi_\mathrm{resp}=0$. This assumption is reasonable if 
\begin{equation}
\lvert\delta n\pi\hbar^2/m\rvert\ll\lvert e\delta\phi_\mathrm{ext}\rvert.\label{condit}
\end{equation}
To understand this condition, consider the influence of a point charge $q$ at the distance $r$ from 2DEG. Then the induced $\delta n$ is of order of $q/(er^2)$, and $\delta\phi_\mathrm{ext}\approx q/(4\pi\varepsilon\varepsilon_0r)$.  The substitution of these expressions into Eq. (\ref{condit}) shows its equivalence to condition $r\gg a_B$, where $a_B\approx 13\ \mathrm{nm}$ is effective Bohr radius in GaAs, which is much less then membrane thickness $t=166\ \mathrm{nm}$. Thus, we believe that the model of pure electrostatic screening allows us to estimate the influence of most of the bound charge induced by a cantilever bending.

To simplify the calculations, we consider the system as an infinite non-bent equipotential plane (2DEG) sandwiched between two $t/2$-thick layers of a material with dielectric constant $\varepsilon\approx 13$ equal to that of $\mathrm{Al_{0.33}Ga_{0.67}As}$. We neglect the small bending in electrostatic calculations, but save the bound charge determined by (\ref{vcharge}) and (\ref{scharge}) at $0\leq l\leq L$ and put it zero otherwise. Using these simplifications, we can estimate the screening charge density using the method of images as follows:
\begin{equation}
\delta n=-\frac{1}{2\pi e}\int_0^L\rho_\mathrm{B}\left[f_\mathrm{surf}\left(\frac{l-l^\prime}{t}\right)+f_\mathrm{vol}\left(\frac{l-l^\prime}{t}\right)\right]dl^\prime,\label{dens}
\end{equation}
where
\begin{flalign}
f&_\mathrm{surf}(r)=\frac{2\varepsilon}{\varepsilon+1}\sum_{n=0}^\infty\left(-\frac{\varepsilon-1}{\varepsilon+1}\right)^n\frac{n+1/2}{r^2+(n+1/2)^2},\\
f&_\mathrm{vol}(r)=-\ln{\left( 1+\frac{1}{2r}\right)}\nonumber\\
&-\sum_{n=1}^\infty\ln{\frac{[r^2+(n+1/2)^2][r^2+(n-1/2)^2]}{[r^2+n^2]^2}}.\label{horror}
\end{flalign}

The calculated $\delta n$ per unit displacement of the cantilever free end $\xi$ is shown in Fig. \ref{fig3}(b) as a function of distance $l$ from the cantilever base. The obtained dependence shows that the considered effect is expected to be most prominent at distances $\lvert l\rvert$ of the order of the cantilever thickness $t$ from its base, with $\delta n$ changing the sign when $l$ is negated. Almost the same dependence can be obtained for the mechanical stress in the form of $\sigma(l=0)\Theta(l)$, where $\sigma$ is determined by (\ref{sigma}) and $\Theta(l)$ is the Heaviside step function. This shows that, in spite of the mechanical stress $\sigma(l)$ lasting to the cantilever free end, the change in electron density $\delta n(l)$ is primarily an edge effect arising near the point $l=0$, where the stress suffers a jump. This point corresponds to the lateral boundary between suspended and non-suspended areas. It is clear that actually $\sigma(l)$ decreases rapidly in the non-suspended bulk at the characteristic distance of order $t$ from the boundary. Thus, we believe that the stepwise jump of the mechanical stress implied by our model can be used as a reasonable simplification.

The distance between the conductive constriction center and the boundary between the suspended and non-suspended areas is measured using a scanning electron microscope and equals $l=1.3\ \mathrm{\mu m}$. The electron density change corresponding to this distance can be obtained from (\ref{dens}) and equals $d(\delta n)/d\xi=7.4\times 10^8\ \mathrm{cm^{-2}\mu m^{-1}}$. Assume that the conductance change  is caused solely by the change in electron density. Then the expected relative conductance sensitivity to the vibrations is equal to the expected relative change in the electron density: $(1/G_0)(dG/d\xi)=(1/n)(d(\delta n)/d\xi)\approx 1.1\times 10^{-3}\ \mathrm{\mu m^{-1}}$. This value agrees with the values $0.56\times 10^{-3}\ \mathrm{\mu m^{-1}}$ and $-0.86\times 10^{-3}\ \mathrm{\mu m^{-1}}$ experimentally obtained above for the Cantilever-I and Cantilever-II. Thus, the proposed model agrees with the experiment and seems to describe adequately the 2DEG conductivity response to mechanical vibrations of the cantilevers, though a detailed experimental study of the spatial change in the electron density near the boundary between suspended and non-suspended areas is desirable.

To conclude, it is experimentally shown that the conductance change resulting from mechanical vibrations of NEMS with 2DEG demonstrates the anisotropy inherent to piezoelectric effect. A model describing this change and predicting its value is proposed. The model implies that the mechanical stress induces the bound charge which is screened by the change in the density of electron gas. According to the model, the change in 2DEG conductivity is related primarily to the rapid change in the mechanical stress near the boundary between suspended and non-suspended areas, rather than to the stress itself.
\begin{acknowledgments}
We thank Tektronix corporation, Alpha-instruments company and Ilya Dianov for providing an AFG3252C arbitraty form generator.\\
The work is supported by RFBR grants 16-32-60130, 15-02-05774 and 16-02-00579.
\end{acknowledgments}
%\nocite{*}

\bibliography{Shevyrin_bibliography}% Produces the bibliography via BibTeX.

\end{document}